\documentclass[12pt]{article}

\newcommand{\x}{arXiv:}
\newcommand{\m}{\mathrm}
\newcommand{\be}{\begin{equation}}
\newcommand{\ee}{\end{equation}}
\newcommand{\ba}{\begin{eqnarray}}
\newcommand{\ea}{\end{eqnarray}}

\usepackage{graphicx}
\usepackage{amssymb}
\usepackage{amsmath}
\usepackage[T1]{fontenc} 
\usepackage[ansinew]{inputenc} 
\usepackage[nosort]{cite}
\newcommand{\inbar}{\vrule height1.57ex width.4pt depth0pt}
\newcommand{\SW}{\relax{\hbox{$\ \inbar\kern-.285em{\rm S}$}}}

\begin{document}
\thispagestyle{empty}
\begin{center}

\null \vskip-1truecm \vskip2truecm

{\Large{\bf \textsf{How Does the Quark-Gluon Plasma Know the }}}

\vskip0.5truecm
{\Large{\bf \textsf{Collision Energy? }}}

\vskip1truecm

{\large \textsf{Brett McInnes}}

\vskip0.5truecm

\textsf{\\  National
  University of Singapore}

\textsf{email: matmcinn@nus.edu.sg}\\

\end{center}
\vskip1truecm \centerline{\textsf{ABSTRACT}} \baselineskip=15pt
\medskip

Heavy ion collisions at the LHC facility generate a Quark-Gluon Plasma (QGP) which, for central collisions, has a higher energy density and temperature than the plasma generated in central collisions at the RHIC. But sufficiently peripheral LHC collisions give rise to plasmas which have the \emph{same} energy density and temperature as the ``central'' RHIC plasmas. One might assume that the two versions of the QGP would have very similar properties (for example, with regard to jet quenching), but recent investigations have suggested that \emph{they do not}: the plasma ``knows'' that the overall collision energy is different in the two cases. We argue, using a gauge-gravity analysis, that the strong magnetic fields arising in one case (peripheral collisions), but not the other, may be relevant here. If the residual magnetic field in peripheral LHC plasmas is of the order of at least $eB\,\approx \,5\,m^2_{\pi}$, then the model predicts modifications of the relevant quenching parameter which approach those recently reported.

\newpage
\addtocounter{section}{1}
\section* {\large{\textsf{1. Peripheral LHC vs. Central RHIC}}}
The novel form of matter produced in heavy ion collisions, the Quark-Gluon Plasma or QGP, is currently under study at two major facilities: the RHIC \cite{kn:RHIC}, which typically collides gold nuclei at maximal centre-of-mass energies up to 200 GeV per nucleon pair, and the LHC, which (in the heavy ion runs studied particularly in ALICE) typically collides lead nuclei at maximal centre-of-mass energies around 2.76 TeV per pair, latterly upgraded to 5.02 TeV \cite{kn:malg1,kn:malg2}. On this basis, it is often said that the two facilities explore two different regimes for the QGP.

There is one case, however, in which the RHIC and the LHC do probe the same regime of temperatures and energy densities. The density of a nucleus prior to a collision is not constant along a transverse direction, since the nucleus tapers away from its central axis. This simple observation has many crucial ramifications, as was pointed out by Becattini et al. \cite{kn:bec}; \emph{inter alia} it implies that the densities and temperatures arising in peripheral collisions at sufficiently high impact parameter are arbitrarily lower than those in central collisions. It follows \cite{kn:hogy,kn:quench,kn:quenchagain,kn:quenchyetagain} that sufficiently \emph{peripheral} collisions studied \cite{kn:centrality1,kn:centrality2} at the LHC give rise to plasmas having the same energy density and temperature as plasmas produced in \emph{central} collisions at the RHIC.

This opens the way to an investigation of a fundamental question: \emph{does the plasma produced in peripheral collisions ``know'' about global parameters like the overall impact energy, or is it sensitive only to explicitly local parameters such as the energy density}?

One way to approach such questions is through studying \emph{jet quenching} (see for example \cite{kn:jet,kn:jetagain} for reviews), the effect of the plasma on the propagation of highly energetic partons produced by the collision. This is conventionally described by the parameter $\hat{q}$, the mean squared transverse momentum acquired by a hard parton per unit distance travelled. In \cite{kn:quench,kn:quenchagain,kn:quenchyetagain} this is represented by a dimensionless parameter $K$, defined by
\begin{equation}\label{A}
K\;\equiv\;{\hat{q}\over 2\epsilon^{3/4}},
\end{equation}
where $\epsilon$ denotes the local energy density; and it is found, in the models constructed there (see also \cite{kn:shanshan}), that this quantity is larger, by a factor\footnote{This is for collisions at 2.76 TeV; values about 15$\%$ larger are quoted \cite{kn:quench} for the 5.02 TeV collisions.} of about 2 to 3, for the RHIC plasmas than for their LHC counterparts. \emph{This is true even when one compares peripheral LHC collisions with central RHIC collisions}, so that, as explained above, the resulting plasmas can have the same energy density and temperature. Thus we have a strong suggestion that the local properties of the ``peripheral'' plasma are, in some way, (very) sensitive to a global parameter, the centre-of-mass energy of the overall collision. This strange and potentially very important development is one aspect of what has been called \cite{kn:JET} the \emph{JET puzzle}.

Work on explanations of this puzzle has begun \cite{kn:cujet,kn:bianchi,kn:kumar}. Here we wish to consider one potentially crucial aspect of the situation that remains to be taken into account: central RHIC plasmas and their peripheral LHC counterparts differ in the following sense: \emph{the latter (only) are immersed in strong magnetic fields}, a fact that has recently given rise to a large literature (see for example \cite{kn:magnet}). This may be relevant, because a strong magnetic field tends to suppress momentum diffusion in directions perpendicular to it \cite{kn:kitu}, and possibly because it subjects the plasma to ``paramagnetic squeezing'' \cite{kn:bali}, altering the pressure gradients; in both cases the (kinematic) \emph{viscosity} of the plasma will be affected. This in turn will affect jet quenching.

Unfortunately, it is very difficult to assess the precise magnitude of the magnetic fields relevant to this situation, and the question is currently under intense scrutiny. The \emph{maximal} possible values of $eB$, those attained in the initial state of the fireball generated by a peripheral collision, were estimated in the classic work of Deng et al. \cite{kn:denghuang}, but it has recently been suggested that these values may need to be reconsidered: in \cite{kn:holliday,kn:peroutka}, for example, it is argued that quantum diffusion effects can significantly adjust the estimated initial magnetic field \emph{upward}.

In addition to these uncertainties, it is difficult to estimate how the field evolves as time passes; this is of course necessary because one needs to use the value of the field at the time when quenching actually occurs. This uncertainty as to the value of the magnetic field at relevant times is one of the most pressing problems for the entire subject: for example, it obviously affects the many investigations regarding the \emph{chiral magnetic effect} \cite{kn:taskforce}. The most elementary observation, that the field must decrease simply due to the departure of the spectator nucleons, has to be tempered by subtle effects connected with the conductivity of the system formed by the collision, penetration depth effects, the interaction of the magnetic field with the associated vorticity, corrections required by event-by-event analyses \cite{kn:bzdak}, and so on: see \cite{kn:gursoy,kn:shipu,kn:inghirami,kn:arpan,kn:dash,kn:shipu2,kn:shub} for balanced discussions of these rather formidable complexities, and the relevant references.

In view of these uncertainties, we will proceed cautiously. Guided by the discussions in \cite{kn:gursoy,kn:shipu,kn:inghirami,kn:arpan,kn:dash,kn:shipu2,kn:shub}, we will consider three scenarios: one in which we directly use the values given in \cite{kn:denghuang}, then one in which those values are decreased, due to the time evolution of the magnetic field, by a factor of 10, and finally one in which the attenuation factor is 100. Note that, even in this last (``most pessimistic'') case, the magnetic fields involved are still enormous (on the order of 10$^{18}$ gauss), so it is far from obvious that there will be no effect even in this case. On the other hand, the viscosity of fluids typically varies \emph{extremely} slowly with pressure, so it is equally unclear that there will be any effect at all, even in the ``most optimistic'' case\footnote{One should also note that there is considerable purely theoretical interest, particularly in lattice theory, in the effect of magnetic fields on the QGP, in the case where $eB$ is allowed to become arbitrarily (that is, not necessarily realistically) large: for a discussion of this interesting line of research, see \cite{kn:gergely}. In this work, however, we do not explore values of the field beyond those given in \cite{kn:denghuang}.}.

The specific question we consider here is whether the $K$ parameter is modified to any \emph{significant} extent by magnetic fields of these orders. In particular, we wish to use a simple gauge-gravity \cite{kn:nat} model to assess whether the magnetic fields in this case, huge though they may be, can give rise to changes in $K$ of the same order of magnitude as those reported in \cite{kn:quench,kn:quenchagain,kn:quenchyetagain}.

Gauge-gravity models of jet quenching are of course well known \cite{kn:hong1,kn:hong2}, and such models also exist which take into account the effect of the magnetic field \cite{kn:mamoquench}. Here we will use a much more basic ``minimal'' model \cite{kn:89,kn:90} in which the bulk geometry is that of a magnetically charged dilatonic asymptotically AdS thermal black hole\footnote{For technical reasons, it is in fact quite difficult to construct such black holes. It was first achieved by Gao and Zhang in \cite{kn:gz}. See \cite{kn:deg} for the generalization to the charged case and to arbitrary dimensions.}. We stress again that there is no reason to expect, over the relatively narrow range of magnetic field values being considered here, any close agreement with the findings of \cite{kn:quench,kn:quenchagain,kn:quenchyetagain}; indeed it is not clear that the holographically computed values of $K$ will even change in the correct direction (\emph{smaller} for the LHC plasmas, by a factor of around 2 or 3, or perhaps a little more for the 5.02 TeV collisions).

We find that this simple model predicts the following. We consider plasma temperatures corresponding to central RHIC collisions and to suitably peripheral LHC collisions (with the associated magnetic field estimated as above, including attenuation factors 1, 10, 100). Then the LHC $K$ parameter is predicted, for centre-of-mass energy 2.76 TeV, to be, respectively, $\approx \,3.34, 1.55, 1.02$  times smaller than the corresponding RHIC value; for centre-of-mass energy 5.02 TeV, the respective values are $\approx \,4.13, 1.95, 1.06$. In short: if the attenuation is by a factor of around 10, corresponding to residual magnetic fields in the $eB \approx 5 - 10 \,m^2_{\pi}$ range (where $m_{\pi}$ is the conventional pion mass), then the $K$ parameter is predicted to be reduced by roughly the factor proposed in \cite{kn:quench,kn:quenchagain,kn:quenchyetagain}.

We do \emph{not} claim that this observation ``explains'' the puzzling sensitivity of jet quenching to the overall collision energy: holography is not so precise an instrument as that, particularly in the case of a model as simple as the one we use here; nor can one yet be fully confident that an attenuation factor of about 10 is appropriate; furthermore, as the authors of \cite{kn:quench,kn:quenchagain,kn:quenchyetagain} carefully point out, the magnitude of the effect itself is not firmly established. We do however wish to argue that the gauge-gravity duality suggests that \emph{a full solution of the puzzle should take into account the possible effect of the magnetic field on the plasma viscosity.}

We begin by briefly outlining the ``minimal'' gauge-gravity model of this situation; then we state the results of our (numerical) investigation of the resulting equations.

\addtocounter{section}{1}
\section* {\large{\textsf{2. A Gauge-Gravity Model of the QGP in a Strong Magnetic Field}}}
In constructing a gauge-gravity model of the QGP, one must always bear in mind that these models are not \emph{always} ``close'' to the real plasma, with behaviour dictated by QCD. At \emph{extremely} high temperatures, QCD is weakly coupled, but any theory of the kind we are considering here will be scale-invariant: that is, if it is strongly coupled at moderate temperatures, then it will never be weakly coupled. Equally, if the temperature is too low, then QCD will of course be a confining theory, but simple holographic models such as we consider here never confine. The models can only be expected to approximate the real plasma for a range of relatively moderate temperatures, where (for example) deconfinement and Debye screening can be expected to arise in \emph{both} systems. (See \cite{kn:nat} for a particularly clear discussion of these points.)

In the situations we consider here (until Section 5, below, where we will have to return to this issue), we are interested in typical RHIC temperatures, and in LHC plasmas with (by LHC standards) relatively low energy densities, and this is precisely the domain of ``relatively moderate temperatures'' required for the duality to be a reasonable approximation ---$\,$ though, even in this case, it remains an \emph{approximation}.

The model we consider has an action in the bulk, describing the interactions of a gravitational field with the usual Einstein-Hilbert Lagrangian, of a dilaton $\varphi$, and of an electromagnetic field $F_{\mu\nu}$, taking the form
\begin{equation}\label{ALEPH}
S=-\frac{1}{16\pi}\int \m{d}^4x \sqrt{-g} \left[R -2(\nabla \varphi)^2 - V(\varphi) - 4\pi\ell_P^2e^{-2\alpha \varphi}F^2\right],
\end{equation}
where $\ell_P$ is the bulk Planck length, and where the dilaton has a potential given by
\begin{equation}\label{BETH}
V(\varphi)\;=\;{- 1\over 8\pi L^2}{1\over (1+\alpha^2)^2}\left[\alpha^2\left(3\alpha^2-1\right)e^{-2\varphi/\alpha}\,+\,\left(3-\alpha^2\right)
e^{2\alpha \varphi}\,+\,8\alpha^2e^{\alpha \varphi - \left(\varphi/\alpha\right)}\right].
\end{equation}
It was shown in \cite{kn:gz} that it is necessary to take this specific form for the potential in order to make it possible to construct a dilatonic bulk metric which is asymptotically AdS, as the gauge-gravity duality requires.

Specifically, we take the bulk geometry to be that of an asymptotically AdS dilatonic magnetically (not electrically) charged Reissner-Nordstr\"om black hole with a flat event horizon (indicated by a zero superscript). As the event horizon here is a plane ---$\,$ see below for the possibility of compactifying it ---$\,$ the charge and mass of the black hole are formally infinite. To get around this, one uses charge and mass \emph{parameters} $P^*$ and $M^*$ which play the role of charge and mass in the following sense: they fix the \emph{charge per unit horizon area}, given by $P^*/(\ell_Pf(r_h)^2)$, and \emph{the mass per unit horizon area}, given by $M^*/(\ell_P^2f(r_h)^2)$, where $r=r_h$ at the event horizon.

We should clarify at this point that, throughout this work, we use natural (not Planck) units; this is the preferred choice for discussing the physics of the (non-gravitational) field theory on the boundary. Thus $P^*$ and $M^*$ are to be regarded as black hole parameters describing the spacetime geometry in the bulk: for reasons of notational convenience they are both taken to have units of \emph{length}; the \emph{physical} charge and mass per unit horizon area are given, as explained above, by $P^*/(\ell_Pf(r_h)^2)$ (units of inverse area, since $P^*$ has units of length, which is correct since charge is dimensionless in natural units) and $M^*/(\ell_P^2f(r_h)^2)$ (units (length)$^{-3}$, since $M^*$ has units of length, which is correct since mass or energy has units of inverse length in natural units). (The reader who finds this confusing should simply remember that \emph{all} bulk parameters, $r_h, f(r)$ (see below), $P^*,$ and $M^*$, have units of length throughout this work.)

The metric takes the form \cite{kn:gz,kn:deg}
\begin{equation}\label{AA}
g(\m{AdSdilP^*RN}^{0})\;=\;-\,U(r)\m{d}t^2 + {\m{d}r^2\over U(r)} + [f(r)]^2 \left[\m{d}\psi^2\;+\;\m{d}\zeta^2\right],
\end{equation}
where $t$ and $r$ are as usual and $\psi$ and $\zeta$ are dimensionless planar bulk coordinates\footnote{If one wishes to compactify in these directions, one can do so; then $\psi$ and $\zeta$ will be angular coordinates on a torus (ranging from $0$ to $2\pi$ in both cases), so we obtain an example of a ``topological'' black hole. This is not necessary here, however: our black hole is topologically trivial, that is, the sections $r = $ constant are true planes, including the event horizon; that is, we take $\psi$ and $\zeta$ to run from $0$ to $\infty$. (Such black holes are sometimes called ``black branes''.) This is customary in applications of such black holes to holography, since one usually does not wish to compactify the space on which the dual field theory propagates.}. The metric coefficients are
\begin{equation}\label{B}
U(r)=-\frac{8\pi M^*}{r}\left[1-\frac{(1+\alpha^2)P^{*2}}{2M^*r}\right]^{\frac{1-\alpha^2}{1+\alpha^2}} + \frac{r^2}{L^2} \left[1-\frac{(1+\alpha^2)P^{*2}}{2M^*r}\right]^{\frac{2\alpha^2}{1+\alpha^2}},
\end{equation}
and
\begin{equation}\label{C}
f(r)^2 = r^2\left(1-\frac{(1+\alpha^2)P^{*2}}{2M^*r}\right)^{\frac{2\alpha^2}{1+\alpha^2}};
\end{equation}
here $\alpha$ is the coupling of the dilaton $\varphi$ to the magnetic field, the corresponding term in the Lagrangian being $\ell_P^2e^{-2\alpha \varphi}F^2$, and $L$ is the AdS curvature scale.

The geometry at infinity is obtained in the usual manner, by factoring out $U(r)$ and allowing r to tend to infinity. Doing this, one finds that the conformal structure at infinity is represented by the metric
\begin{equation}\label{AAA}
g(\m{AdSdilP^*RN}^{0})_{\infty}\;=\;-\,\m{d}t^2 + L^2 \left[\m{d}\psi^2\;+\;\m{d}\zeta^2\right];
\end{equation}
setting $x = L\psi$ and $z = L\zeta$, and adjoining a third Cartesian coordinate $y$, one has flat spacetime at infinity, with a length scale set by $L$. This length scale must be chosen on physical grounds: the most natural choice is to take it to be the characteristic length scale of the QGP fireball generated by the collision, say $L \approx$ 10 fm. We shall use this value here. Note that, as is customary in nuclear physics, we are focussing on the \emph{reaction plane}, the plane described by the coordinate $z$ (along the axis of the collision) and $x$ (transverse to $z$). Holography describes the physics in this plane; we slice the full three-dimensional collision region into slices of the form $y = $ constant, and study each slice independently.

For the purposes of this work, the reader may regard the presence of the dilaton as essentially a technical matter, required (when magnetic fields are extremely strong relative to the squared temperature) in order to ensure that the bulk system is mathematically consistent within string theory\footnote{The theory in the bulk is ultimately a string theory, and the internal consistency of string theory imposes many conditions on the bulk parameters: see the discussions in \cite{kn:ferrari1,kn:ferrari2,kn:ferrari3,kn:ferrari4} for the full details.}. The form taken by this mathematical consistency condition in the present application is discussed in detail in \cite{kn:ong,kn:89,kn:90}. The value of the coupling $\alpha$ is chosen to be the minimal value that results in this condition being satisfied. These considerations fix the value of $\alpha$ when the temperature and magnetic charge of the black hole are given: see below for a brief discussion of how this works in practice.

The conformal transformation used above to relate the bulk and boundary geometries has the effect of imprinting the bulk magnetic field on conformal infinity: this is a procedure of basic importance in applications of gauge-gravity duality to condensed matter theory \cite{kn:hartkov}. In the present case it works as follows: the electromagnetic field two-form here takes the form
\begin{equation}\label{AAAA}
F\;=\;{P^*\over \ell_P}\left(1-\frac{(1+\alpha^2)P^{*2}}{2M^*r}\right)^{\frac{2\alpha^2}{1+\alpha^2}}\,\m{d}\psi \wedge \m{d}\zeta,
\end{equation}
and letting $r$ tend to infinity (and recalling that $x = L\psi$ and $z = L\zeta$), we find, after fixing the scaling freedom\footnote{When one uses a conformal transformation to determine the geometry at infinity for an asymptotically AdS black hole, the curvature at infinity is related to the curvature of the event horizon. But when the latter is zero, as it is here, there is an overall scaling freedom, which can be exploited to perform a field redefinition which replaces $\ell_P$ in this formula with $L$. (Recall that the value of the magnetic field itself is obtained by using an \emph{orthonormal} basis in this formula, so the scaling of the metric has an effect here.) This is appropriate, since $L$ sets the scale of the (non-gravitational) boundary theory, whereas $\ell_P$ is of course an intrinsically gravitational scale. This procedure was first employed (without discussion) in \cite{kn:hartkov}; recently it has been discussed in detail in
section 2.2.1 of \cite{kn:newmyers}.},
\begin{equation}\label{D}
B_{\infty}\;=\; {P^*\over L^3}.
\end{equation}
As in \cite{kn:hartkov}, this magnetic field is parallel to the $y$ axis and is completely uniform within the reaction plane, so the holographic picture does not lead to unphysical behaviour, such as monopoles. (The actual net magnetic field in the QGP does indeed point, on average, along the $y$ axis, perpendicular to the reaction plane; it is independent of position in that plane to a good approximation \cite{kn:ferrer}.)

The equation for $r_h$ is just, from (\ref{B}),
\begin{equation}\label{E}
-\frac{8\pi M^*}{r_h}\left[1-\frac{(1+\alpha^2)P^{*2}}{2M^*r_h}\right]^{\frac{1-\alpha^2}{1+\alpha^2}} + \frac{r_h^2}{L^2} \left[1-\frac{(1+\alpha^2)P^{*2}}{2M^*r_h}\right]^{\frac{2\alpha^2}{1+\alpha^2}} = 0,
\end{equation}
and the Hawking temperature of this black hole, corresponding to the temperature at infinity, is
\begin{eqnarray}\label{F}
4\pi T_{\infty}&=&{8\pi M^*\over r_h^2}\left(1-\frac{(1+\alpha^2)P^{*2}}{2M^*r_h}\right)^{{1-\alpha^2\over 1+\alpha^2}}\;-\;{4\pi (1-\alpha^2)P^{*2}\over r_h^3}\left(1-\frac{(1+\alpha^2)P^{*2}}{2M^*r_h}\right)^{{-2\alpha^2\over 1+\alpha^2}}\;\nonumber \\ &
&+\;{2r_h\over L^2}\left(1-\frac{(1+\alpha^2)P^{*2}}{2M^*r_h}\right)^{{2\alpha^2\over 1+\alpha^2}}\;+\;{\alpha^2P^{*2}\over M^*L^2}\left(1-\frac{(1+\alpha^2)P^{*2}}{2M^*r_h}\right)^{{\alpha^2 - 1 \over 1+\alpha^2}}.
\end{eqnarray}

Given $B_{\infty}$, $T_{\infty}$ and $L$ (and therefore $\alpha$), one can use the three equations (\ref{D}), (\ref{E}), and (\ref{F}) to solve (in physical cases) for $r_h$, $P^*$, and $M^*$, and in this way the ``known'' boundary parameters fix the bulk geometry. In particular, therefore, they determine (together with the bulk Planck length) the black hole entropy per unit horizon area and the black hole mass per unit horizon area. The former is $1/4\ell_P^2$ for Einstein gravity, which is all we use in this minimal model; the latter is given, as above, by $M^*/(\ell_P^2f(r_h)^2)$. The ratio of these quantities is dual to the ratio of the entropy density of the boundary system, $s$, to its energy density $\epsilon$, so we have
\begin{equation}\label{G}
{s\over \epsilon}\;=\;{f(r_h)^2\over 4M^*} \;=\; {r_h^2\left(1-\frac{(1+\alpha^2)P^{*2}}{2M^*r_h}\right)^{\frac{2\alpha^2}{1+\alpha^2}}\over 4 M^*}.
\end{equation}
Given $B_{\infty}$, $T_{\infty}$, and $\epsilon$ for the boundary theory, we can use this to make a holographic prediction regarding $s/\epsilon$. (Note that $s$ has units of fm$^{-3}$ in natural units, $\epsilon$ has units of fm$^{-4}$, so the left side of (\ref{G}) has units of length; bearing in mind that all bulk quantities have units of length here, this agrees with the right side.)

This is relevant here because the quenching parameter $\hat{q}$ is closely related to the entropy density. In fact \cite{kn:hong1,kn:hong2}, in a strictly conformal, strongly coupled boundary theory, one expects $\hat{q}$ to scale with the square root of $s$. (Of course, the real plasma does not correspond to a conformal theory; the errors thus inevitably introduced can however be estimated \cite{kn:hong1,kn:hong2}, and it appears that they will not invalidate the kind of rough estimates we are aiming for here.)

We are now in a position to compare $K^C_{RHIC}$, the $K$-parameter for central RHIC collisions, with $K^P_{LHC}$, the value for peripheral LHC collisions resulting in plasmas at the same temperature.

\addtocounter{section}{1}
\section* {\large{\textsf{3. A Holographic Computation of $K^C_{RHIC}/K^P_{LHC} (\sqrt{s_{NN}} = 2.76 \;TeV)$}}}
We are interested in comparing the plasmas produced in central Au-Au RHIC collisions (at a temperature \cite{kn:phobos} of around 220 MeV after equilibration) with those produced in Pb-Pb LHC collisions (at a centre-of-mass collision energy around 2.76 TeV per pair) which are sufficiently peripheral as to give rise to the same temperature and energy density.

For the central RHIC collisions we have $B = \alpha = 0$; it is now easy to solve (\ref{E}) and (\ref{F}) for $r_h$ and $M^*$, and then $s^C_{RHIC}/\epsilon$ can be found from equation (\ref{G}).

The energy density for \emph{central} LHC collisions is estimated \cite{kn:aliceenergy} to be around 2.3 times as large as in central RHIC collisions; using this, and following the discussion of the ``thickness function'' for nuclei given in \cite{kn:bec}, we find that the LHC plasmas in which we are interested arise when the impact parameter of the collision is $b\approx 12$ fm (with an assumed nuclear radius of around 7 fm). Consulting \cite{kn:denghuang} one finds that this corresponds to an initial magnetic field of about $eB_{\infty} \approx 60\,m^2_{\pi}$, where $m_{\pi}$ is the standard pion mass. As explained above, this means that we will consider three values for $eB_{\infty}$: $eB_{\infty} \approx 60\,m^2_{\pi}$, $eB_{\infty} \approx 6\,m^2_{\pi}$, and $eB_{\infty} \approx 0.6\,m^2_{\pi}$. These data allow us to compute $\alpha \approx 0.34, \alpha \approx 0, \alpha \approx 0$ for the three respective cases\footnote{The fact that we can take $\alpha$ to be approximately zero in the second and third cases can be understood in terms of the inequality discussed in \cite{kn:88}: we saw there that the inconsistency problem arises (in the absence of any other mitigating effect, such as the dilaton or vorticity) only if the inequality
\begin{equation}\label{GG}
B_{\infty}\;\leq \;2\pi^{3/2}T^2\;\approx \; 11.14 \times T_{\infty}^2
\end{equation}
is violated. For the temperature we are considering here, this begins to happen only when $eB_{\infty}$ reaches $\approx 8.37\,m^2_{\pi}$, so the dilaton is not needed for $eB_{\infty} \approx 6\,m^2_{\pi}$ or $eB_{\infty} \approx 0.6\,m^2_{\pi}$.}, after the manner of \cite{kn:89,kn:90}. We can now solve (\ref{D}), (\ref{E}), and (\ref{F}) for $P^*$, $r_h$, and $M^*$ in each of the three cases, and again equation (\ref{G}) yields three values for $s^P_{LHC}/\epsilon$; note that, by construction of course, $\epsilon$ is the same throughout these computations.

We can now proceed in a straightforward way: we have, in an obvious notation,
\begin{equation}\label{H}
 {K^C_{RHIC}\over K^P_{LHC}}\;=\;{\hat{q}^C_{RHIC}/2\epsilon^{3/4}\over \hat{q}^P_{LHC}/2\epsilon^{3/4}}\;=\;{\hat{q}^C_{RHIC}\over \hat{q}^P_{LHC}}\;=\;\sqrt{{s^C_{RHIC}\over s^P_{LHC}}}\;=\;\sqrt{{s^C_{RHIC}/\epsilon\over s^P_{LHC}/\epsilon}},
\end{equation}
and this last quantity is something we can, as explained, compute holographically. A simple numerical analysis of the equations discussed above yields the following results:
\begin{equation}\label{I60}
 {K^C_{RHIC}\over K^P_{LHC}(\sqrt{s_{NN}} = 2.76 \;TeV)}(eB\,=\,60\,m^2_{\pi})\;\approx \; 3.34.
\end{equation}
\begin{equation}\label{I6}
 {K^C_{RHIC}\over K^P_{LHC}(\sqrt{s_{NN}} = 2.76 \;TeV)}(eB\,=\,6\,m^2_{\pi})\;\approx \; 1.55.
\end{equation}
\begin{equation}\label{I0.6}
 {K^C_{RHIC}\over K^P_{LHC}(\sqrt{s_{NN}} = 2.76 \;TeV)}(eB\,=\,0.6\,m^2_{\pi})\;\approx \; 1.02.
\end{equation}

We see that, in the ``reasonably optimistic'' case in which the residual magnetic field is around $eB\,=\,6\,m^2_{\pi}$, this indicates that the field does reduce the value of $K$ to a significant extent, though perhaps the effect is not quite as large as the one reported in \cite{kn:quench,kn:quenchagain,kn:quenchyetagain}.

\addtocounter{section}{1}
\section* {\large{\textsf{4. A Holographic Prediction for $K^C_{RHIC}/K^P_{LHC} (\sqrt{s_{NN}} = 5.02 \;TeV)$}}}
The calculations in \cite{kn:quench,kn:quenchagain,kn:quenchyetagain} are primarily concerned with LHC collisions at 2.76 TeV per pair. The holographic technique is easily extended to the recent runs \cite{kn:5.02,kn:latest5.02data} at 5.02 TeV per pair, as follows.

The energy density for central collisions at very high collision energies is discussed, for example, in \cite{kn:FCC2}. We will assume that a typical energy density for central Pb-Pb LHC collisions at 5.02 TeV per pair is around 3 times the maximal RHIC density. Using this in the same manner as above, we find that the impact parameter of the relevant collisions (resulting in a plasma with the same temperature, about 220 MeV, as plasmas produced in central RHIC collisions) is slightly larger than before, $b\approx 13$ fm. Again consulting \cite{kn:denghuang} (noting in particular that the magnetic field at a given location increases roughly linearly with the impact energy), one finds that now the initial magnetic field satisfies $eB_{\infty} \approx 120\,m^2_{\pi}$, so, as before, we consider also $eB_{\infty} \approx 12\,m^2_{\pi}$ and $eB_{\infty} \approx 1.2\,m^2_{\pi}$. We compute $\alpha \approx 0.36$, $\alpha \approx 0.13$, and $\alpha = 0$ respectively with these data. Solving (\ref{D}), (\ref{E}), and (\ref{F}) in these cases and using (\ref{G}), we compute values for $s^P_{LHC}/\epsilon$ which are still smaller than the value for 2.76 TeV collisions, and the final results (using (\ref{H}) again) are as follows:
\begin{equation}\label{K120}
 {K^C_{RHIC}\over K^P_{LHC}(\sqrt{s_{NN}} = 5.02 \;TeV)}(eB\,=\,120\,m^2_{\pi})\;\approx \; 4.13.
\end{equation}
\begin{equation}\label{K12}
 {K^C_{RHIC}\over K^P_{LHC}(\sqrt{s_{NN}} = 5.02 \;TeV)}(eB\,=\,12\,m^2_{\pi})\;\approx \; 1.95.
\end{equation}
\begin{equation}\label{K1.2}
 {K^C_{RHIC}\over K^P_{LHC}(\sqrt{s_{NN}} = 5.02 \;TeV)}(eB\,=\,1.2\,m^2_{\pi})\;\approx \; 1.06.
\end{equation}
Notice that the prediction here, in the case of a 10-fold attenuation of the field, is in somewhat better agreement with \cite{kn:quench} (where values roughly 15$\%$ larger than in the 2.76 TeV case are suggested) than the corresponding prediction for 2.76 TeV collisions; though it is still somewhat low.

We conclude that the holographic approach indicates that the local plasma in the peripheral case ``knows'' about the global impact energy \emph{at least partly} through the magnetic field (though one should also bear in mind the effects discussed in \cite{kn:cujet,kn:bianchi,kn:kumar}).

We suggest a physical interpretation of this result as follows. Note that the relevant property of the plasma here is its ``momentum diffusivity'' or kinematic viscosity $\nu$, defined as the ratio of the dynamic viscosity $\eta$ to the energy density. Then we have
\begin{equation}\label{J}
\nu\;=\;{\eta\over \epsilon}\;=\;{\eta\over s}\times{s\over \epsilon}\;=\;{1\over 4\pi}\times {s\over \epsilon},
\end{equation}
where we have used the well-known KSS relation\footnote{The KSS relation holds as long as the gravitational action is that of ordinary Einstein gravity coupled to matter of any form: see \cite{kn:nat} for a discussion of this remarkable result. It can be violated if one uses higher-derivative corrections to the Einstein action, but we are not doing so here (see equation (\ref{ALEPH}) above).} \cite{kn:KSS} in the last step. (Of course, for the actual plasma one would replace $1/4\pi$ by a somewhat larger (mildly temperature-dependent) quantity; see for example \cite{kn:QGPparameters}. This does not matter here since the temperature is fixed throughout our discussions.) From this we see that the kinematic viscosity is, according to holography, affected by the magnetic field in just the same way as $s/\epsilon$: that is, these extremely intense fields reduce the momentum diffusion to a significant extent \cite{kn:90}. This can be expected to influence the jet quenching parameter, due to the effects mentioned earlier.

\addtocounter{section}{1}
\section* {\large{\textsf{5. Is the Effect Independent of Centrality?}}}
In \cite{kn:quench,kn:quenchagain,kn:quenchyetagain} it is claimed that the value of the $K$-parameter in collisions at a given impact energy is (of course, approximately) \emph{independent} of the centrality of the collision. (The effect is particularly striking for LHC collisions, so we focus on those.) This presents a challenge for the ``magnetic'' theory, because, for a given impact energy, the magnetic fields are very much smaller at small values of $b$ than at large values of $b$ \cite{kn:denghuang}. Is it possible that such different magnetic fields could have much the same effect on $K$? Granted that, as we have argued, high values of $eB$ tend to depress $K$, would one not expect that $K$, as a function of $b$, should show a definite downward trend?

We have computed (in the same manner as above) the ratio of the holographically predicted value of $K$ for the LHC plasmas produced in peripheral collisions at 2.76 TeV per pair, as a function of the impact parameter $b$ (denoted $K^P_{LHC}(b)$), to its value for central LHC collisions (denoted $K^C_{LHC}$), also at 2.76 TeV. The results are shown in Figure 1.
\begin{figure}[!h]
\centering
\includegraphics[width=1\textwidth]{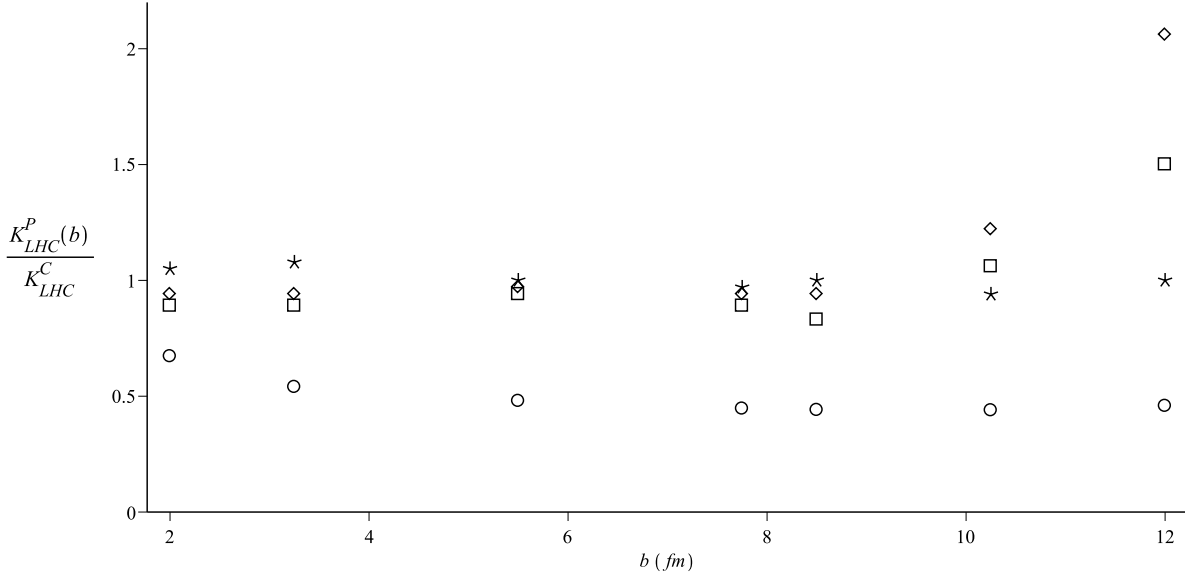}
\caption{Holographic prediction for $K^P_{LHC}(b)/K^C_{LHC}$, for collisions at 2.76 TeV per pair, represented by circles; compared with data taken (with permission) from \cite{kn:quenchagain}, using various hydrodynamic profiles (asterisk = Hirano, box = fKLN, diamond = Glauber), and the free-streaming extrapolation for times prior to thermalization; for the error bars, see \cite{kn:quenchagain}.}
\end{figure}

If we compare this with the data presented in Figure 2 in \cite{kn:quenchagain}, which we have used in Figure 1 to present the corresponding $K^P_{LHC}(b)/K^C_{LHC}$ values (by taking $K^C_{LHC}$ to correspond to the values at the smallest value of $b$), it must be admitted that the predicted points are not in good agreement with any of the three models used there (Hirano, fKLN, Glauber models; note that, in this case, the free-streaming extrapolation has been used for the pre-thermalization physics). (For a more recent discussion of the behaviour at very large $b$, see \cite{kn:quenchyetagain}.)

It is clear, then, that work remains to be done before we can claim that this simple model can describe the results of \cite{kn:quench,kn:quenchagain,kn:quenchyetagain}. We are however encouraged by the following observation: from \cite{kn:denghuang}, the increase in the magnetic field in passing from $b \approx 2$ fm to $b \approx 12$ fm is very large: from $eB \approx 10m_{\pi}^2$ to a colossal $eB \approx 65m_{\pi}^2$. And yet we see that, in the holographic model, this huge increase in the magnetic field has almost no \emph{further} effect on the computed ratio\footnote{In order to make the point clearly, we have neglected the effect of attenuation of the magnetic fields; including that effect would essentially just raise the height of the graph (and, in fact, make it look flatter).}. Thus, while the holographic model is not adequate to give a detailed quantitative account of the variation of $K$ with $b$, it does suggest that the magnetic field does not depress the value of $K$ excessively. This may be the starting point for constructing a more realistic model.

\addtocounter{section}{1}
\section* {\large{\textsf{6. Conclusion: The Magnetic Field May Connect Local with Global Parameters}}}
The puzzling claims \cite{kn:quench,kn:quenchagain,kn:quenchyetagain} that jet quenching can detect the difference between the QGP produced in central RHIC collisions and the plasma produced in peripheral LHC collisions, even when the local temperature and energy densities are the same, and that this effect is independent of the centrality, demand an explanation. We have argued, using a very simple gauge-gravity model, that the extremely intense magnetic fields arising in one case (the LHC plasmas), and not the other, may contribute to an understanding of these claims. The model suggests [a] that the magnetic field, if it is not diluted too strongly, affects jet quenching, possibly by reducing the kinematic viscosity of the plasma, and that [b] this effect nevertheless does not depress the $K$ parameter excessively, even in the case of the huge magnetic fields encountered in very peripheral collisions; though more work is needed if quantitative agreement is sought.

The conclusion is simply that further investigations of these intriguing observations should focus on the effect of the magnetic fields arising in peripheral heavy-ion collisions. In particular it would be useful to have a better understanding of the effect of strong fields on QGP viscosity, focusing perhaps on paramagnetic squeezing \cite{kn:bali}, and using more sophisticated gauge-gravity models than the one employed here.

We close with the following remark. It is suggestive that, in all (realistic) cases considered in Sections 3 and 4, the model leads to somewhat low estimates of the ratio of $K$ in the two kinds of collision. It may be that this means that the magnetic field \emph{alone} cannot explain the full extent of this effect.

There is, however, another, closely related difference between the two cases: in the peripheral case only, huge values of the \emph{vorticity} are produced, an effect long predicted (see for example \cite{kn:liang,kn:bec,kn:huang,kn:viscous,kn:deng,kn:vortical}) and recently observed, by the STAR collaboration \cite{kn:STARcoll}, in the form of global $\Lambda$ hyperon polarization \cite{kn:qunwang}. This could be relevant to the questions raised in this work, either directly or through the subtle interactions of vorticity with magnetic fields \cite{kn:dash,kn:chern}. It remains to be seen whether including this effect can improve the holographic predictions, particularly since the attenuation effect considered here does not apply to that case.

\addtocounter{section}{1}
\section* {\large{\textsf{Acknowledgement}}}
The author is most grateful to Carlota Andrés and the other authors of \cite{kn:quenchagain} for permission to use data from that work.

\end{document}